\documentclass[apjl]{emulateapj}
\usepackage[english]{babel}
\usepackage{graphicx,color}
\usepackage{hyperref}

\newcommand{\FeXII}{\ion{Fe}{12}}

\newcommand{\kms}{km~s$^{-1}$}

\begin{document}

\title{Hi-C Observations of Sunspot Penumbral Bright Dots}


\author{Shane E. Alpert}
\affil{Department of Physics and Astronomy, Rice University, Houston, TX, 77005, USA}
\and
\author{Sanjiv K. Tiwari, Ronald L. Moore, Amy R. Winebarger, and  Sabrina L. Savage}
\affil{NASA Marshall Space Flight Center, ZP 13, Huntsville, AL 35812, USA}


\begin{abstract}
We report observations of bright dots (BDs) in a sunspot penumbra using High Resolution Coronal Imager (Hi-C) data in 193 \AA\ and examine their sizes, lifetimes, speeds, and intensities. The sizes of the BDs are on the order of 1\arcsec\ and are therefore hard to identify in the Atmospheric Imaging Assembly (AIA) 193 \AA\ images, which have 1.2\arcsec\ spatial resolution, but become readily apparent with Hi-C's five times better spatial resolution. We supplement Hi-C data with data from AIA's 193 \AA\ passband to see the complete lifetime of the BDs that appeared before and/or lasted longer than Hi-C's 3-minute observation period. Most Hi-C BDs show clear lateral movement along penumbral striations, toward or away from the sunspot umbra. Single BDs often interact with other BDs, combining to fade away or brighten. The BDs that do not interact with other BDs tend to have smaller displacements. These BDs are about as numerous but move slower on average than Interface Region Imaging Spectrograph (IRIS) BDs, recently reported by \cite{tian14}, and the sizes and lifetimes are on the higher end of the distribution of IRIS BDs. Using additional AIA passbands, we compare the lightcurves of the BDs to test whether the Hi-C BDs have transition region (TR) temperature like that of the IRIS BDs. The lightcurves of most Hi-C BDs peak together in different AIA channels indicating that their temperature is likely in the range of the cooler TR ($1-4\times 10^5$ K). 
\end{abstract}


\section{Introduction}
In the past few decades, high resolution images from space telescopes have allowed solar astronomers to view and analyze the Sun in unprecedented detail. The data from the High Resolution Coronal Imager \citep[Hi-C: ][]{koba14} yield new information about small-scale structures in the corona and chromosphere-corona transition region (TR), including braiding in coronal loops, triggering of subflares in them, and signatures of nanoflares and impulsive heating \citep[e.g.,][]{cirt13,wine13,thal14,tiw14}. This paper examines a sunspot in Hi-C's field-of-view (FOV) with particular attention to the penumbra. A sunspot consists of the inner, darker umbra, and the outer, lighter (but darker than quiet Sun) penumbra. The umbra has strong nearly vertical magnetic field, with umbral-dots and light-bridges in it as signatures of magnetoconvection \cite[e.g.,][]{sobo93,weis02,rimm08,riet13}. The magnetic field inclination in the penumbra greatly varies azimuthally at small-scales \citep{sola03,shibu03,tiw09b,borr11,scha12,tiw13,tiw15aa}. The penumbra is known for its filaments, horizontal tendrils that extend outward from the umbra. Penumbral filaments contain a general outflow \citep[the Evershed flow:][]{ever09}; localized up- and downflows at their heads (ends of filaments closer to sunspot center) and tails, respectively. Penumbral filaments have a stretched granule shape and reversed polarity edge field sustained by lateral downflows \citep{remp12,tiw13}. 


In addition to the dynamic features in the photosphere in sunspots e.g., p-modes, inward motion of penumbral grains, moving magnetic features \citep{harv73}, there exist chromospheric and coronal dynamic features in sunspots e.g., umbral flashes and running penumbral waves \citep{ziri72}, and penumbral jets \citep{kats07,jurc08,tiw16}. There have been only limited investigations of the smaller dynamic events owing to the limited simultaneous temporal and spatial resolution of observations. 

Using data from the Interface Region Imaging Spectrograph \citep[IRIS:][]{depo14}, \cite{tian14} recently discovered bright dots (BDs), another dynamic feature, in the TR of sunspots. The BDs are most numerous in penumbrae but are also present in umbrae and light-bridges. The penumbral BDs observed by IRIS move along the direction of penumbral filaments with speeds of 10--40 \kms, and appear slightly elongated along the filaments, with the two dimensions being 300--600 km and 250--450 km, respectively. The lifetimes of the IRIS BDs are mostly less than one minute, some lasting for a few minutes. Many of these BDs were proposed to be signatures of small-scale energy release events at the TR footpoints of coronal magnetic loops.

In this Paper, we report on Hi-C observations of BDs in a sunspot penumbra. We measure and compare the physical characteristics of these Hi-C penumbral BDs to the IRIS penumbral BDs and use different Atmospheric Imaging Assemply \citep[AIA:][]{leme12} channels to explore the temperature of the Hi-C BDs.

 \section{Observation and Data Analysis} 


\begin{figure*}[tp]
     \centering
     \includegraphics[trim=0.01cm 1.6cm 0.55cm 2.26cm,clip,width=0.88\textwidth]{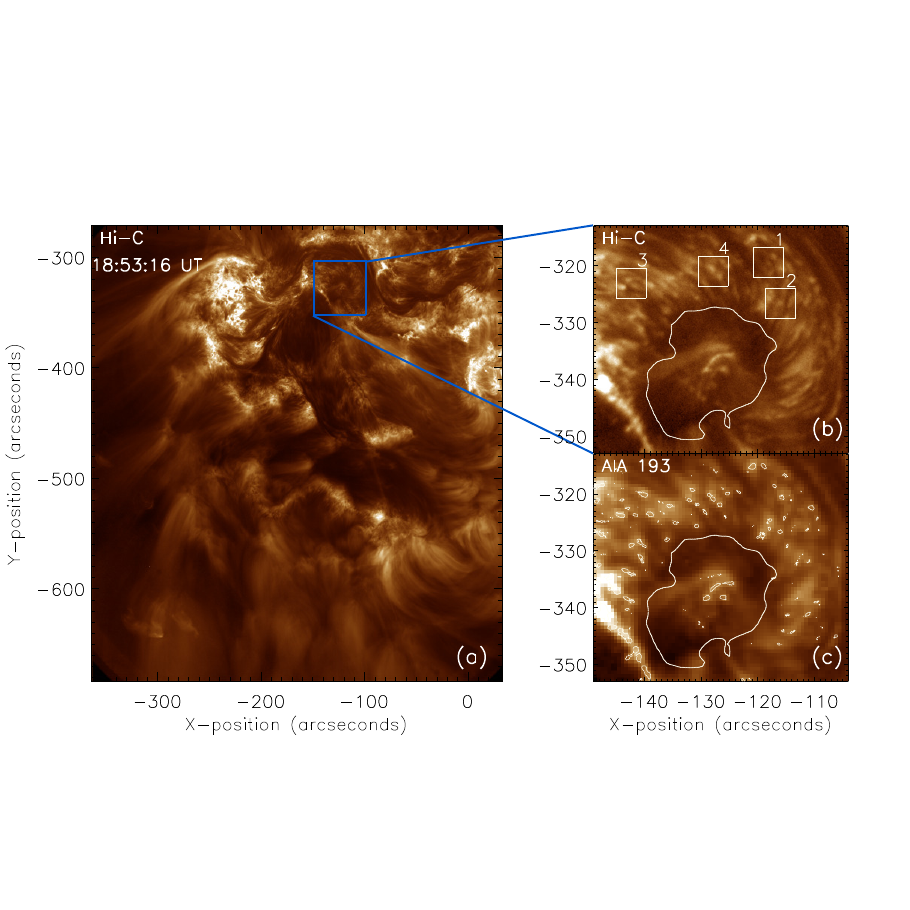}  
     \includegraphics[trim=.02cm .6cm 0.5cm 1.1cm,clip,width=0.8\textwidth]{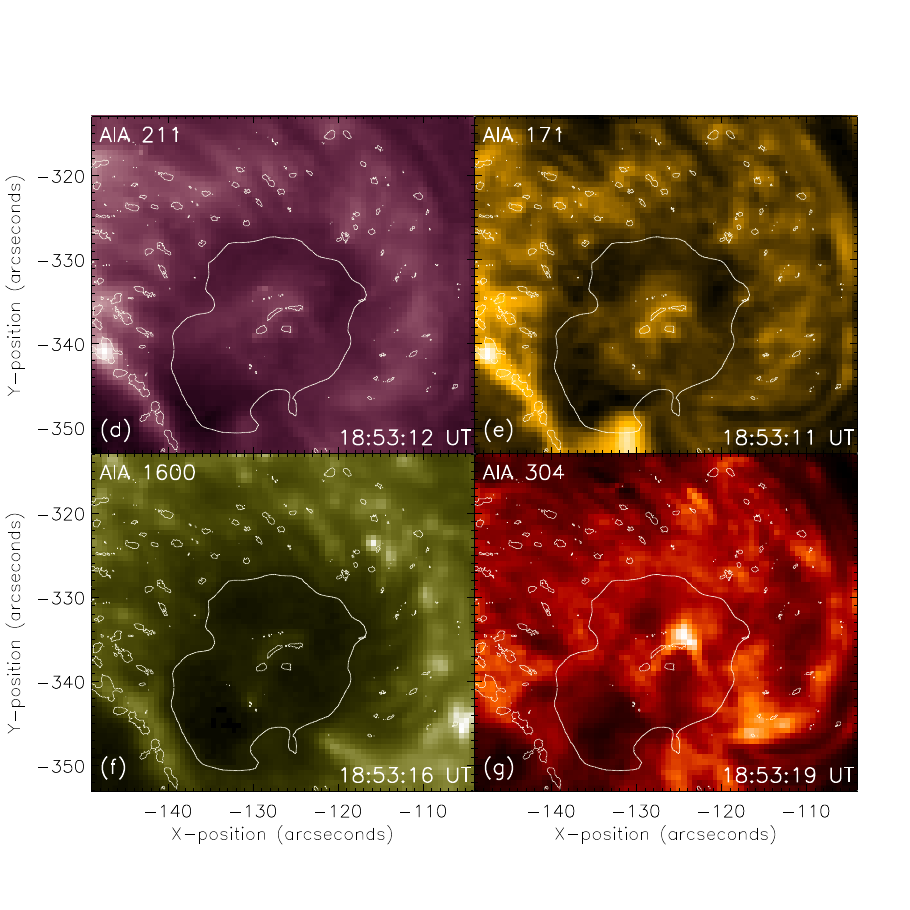}
          \caption{\footnotesize (a) Full FOV of the Hi-C, $\sim$400\arcsec $\times$ 400\arcsec. The blue box covers the sunspot studied here. (b) depicts a close-up view ($\sim$40\arcsec $\times$ 40\arcsec) of the sunspot as seen by Hi-C. The white boxes 1, 2, 3, and 4 outline the locations of example BDs shown in Figure \ref{f2}a, b, c, and d, respectively. (c) is the same FOV and time as (b), but from the AIA 193 \AA\ movie. (d-g) The same 40\arcsec $\times$ 40\arcsec FOV as in (b) and (c), but in other AIA bands. (d), (e), (f) and (g) are 211, 171, 1600 and 304 \AA, respectively. Smaller contours outline Hi-C BDs and are from an enhanced Hi-C image. The large contour roughly outlines the sunspot umbra; most of the outside displayed area being penumbra.}
     \label{f1}
\end{figure*}

We use Hi-C 193 \AA\ data taken on July 11, 2012 during 18:52:43--18:55:30 UT. The entire Hi-C FOV is shown in Figure \ref{f1}a. Hi-C data has a cadence of 5.5 seconds and a plate scale of 0.1\arcsec pixel$^{-1}$. The sunspot that is the main focus of this research is centered south-east of disk center, at ($-$122\arcsec, $-$325\arcsec). A 40\arcsec $\times$ 40\arcsec\ FOV around this point is used for most of the analysis. The blue inset in Figure \ref{f1}a indicates the Hi-C 40\arcsec $\times $ 40\arcsec\ FOV used in this study, enlargement of which is shown in Figure \ref{f1}b. Due to the Hi-C data's short duration and for comparison reasons, supplemental 193 \AA\ data is used from SDO/AIA. AIA has a cadence of 12 seconds (24 seconds for UV channels) and a plate scale of 0.6\arcsec\ pixel$^{-1}$. Figure \ref{f1}c is the same FOV and time of observation as Figure \ref{f1}b, but shows SDO/AIA 193\AA\ data. Hi-C BDs are indicated by the smaller contours. Evidently, BDs are present in both umbra and penumbra of the sunspot, most of them being in penumbra. In this Paper, we have focused on the penumbral BDs.

Additional AIA data is used from the 1600, 304, 171, and 211 \AA\ passbands to examine BD visibility at different atmospheric heights/temperatures. To compare the same FOV across instruments, same subregion of the AIA 193 \AA\ image was extracted to match the Hi-C FOV. These AIA 193 \AA\ coordinates were then used for the other AIA channels alignments.

When examining the sunspot, we find small, roundish, BDs. Any compact transient blob that appears brighter than the surrounding area is considered a BD. To identify, track, and measure the BDs easily, we smooth the images (20 $ \times $ 20 pixels for Hi-C and 7 $\times $ 7 pixels for AIA) and subtract them from the originals. The enhanced images can be seen in comparison to the original images in Movie1. It must be noted that although we used the enhanced images to more easily identify the BDs, all size, intensity, lifetime, speed, and lightcurve measurements were made using the original images. 

\begin{figure*}
     \centering
     \includegraphics[width=\textwidth]{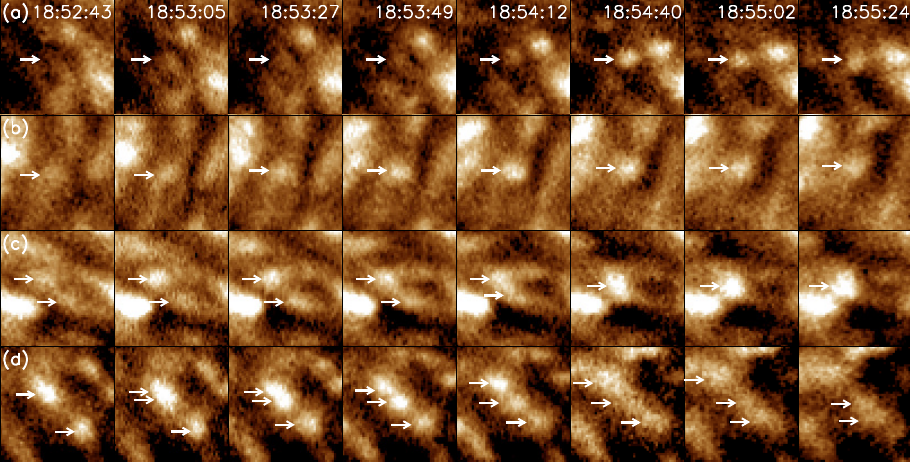}  
     \caption{Examples of four different types of Hi-C penumbral BDs; how they evolve over a three minute period, with arrows pointing to the centers of the BDs. (a) A single BD is stationary while brightening and then fading. (b) A single BD moves laterally outward from sunspot center while brightening and then fading. (c) Two BDs move toward each other and combine into a larger, brighter BD. (d) A system of three BDs. A large BD on the upper left separates into two BDs, with the lower one moving laterally towards sunspot center. A single BD on the lower right moves laterally away from sunspot center. The inward moving BD and the outward moving BD merge as they fade. The size of each panel is 5\arcsec $\times$ 5\arcsec, and their locations are outlined in Figure \ref{f1}(b) by four white boxes numbered 1, 2, 3, and 4.}
     \label{f2}
\end{figure*}

Figure \ref{f2} shows examples of different types of BDs that are seen in the sunspot penumbra. The image sequences display the lateral drift and brightness evolution of the BDs over the Hi-C observing period. Figure \ref{f2}a shows a single stationary BD. Figure \ref{f2}b shows a single BD that moves away from the umbra. Figure \ref{f2}c shows two separate BDs moving towards each other before combining to form one larger, brighter BD. Figure \ref{f2}d shows a triple-system of BDs; a larger BD splits to become two separate, smaller BDs and one of these then combines with a third BD. Unlike the combined, large BDs in Figure \ref{f2}c, the BDs in Figure \ref{f2}d fade as they merge. It must be noted that all BD motion, if motion is seen at all, is along the direction of the penumbral filamentary structures. During its lifetime, a BD may move inward and outward, i.e., oscillatory motion is commonly seen. To look at all of the BDs in the sunspot, see Movie1. 

 We tracked 30 BDs by eye and characterized them. To get the best sense of the movement and overall changes of the BDs, we picked one at a time and tracked throughout the length of the movie, combining Hi-C and AIA data. We repeated this for a few movie cycles, and for each of 30 BDs. Note that for the BDs that separate into smaller BDs, a variety of motion is seen after the split; the individual BDs may move apart, remain stationary, one moves while the other remains stationary, or they may recombine to either form a larger, brighter BD (Figure \ref{f2}c) or fade away (Figure \ref{f2}d). The longest-lived BDs tend to be the ones that are large, bright, and isolated (no interaction with other BDs).

\section{Results and Discussion} 

\begin{figure*}[tp]
     \centering
     \includegraphics[trim=0.36cm 0.55cm 0.56cm 0.7cm, clip, width=\textwidth]{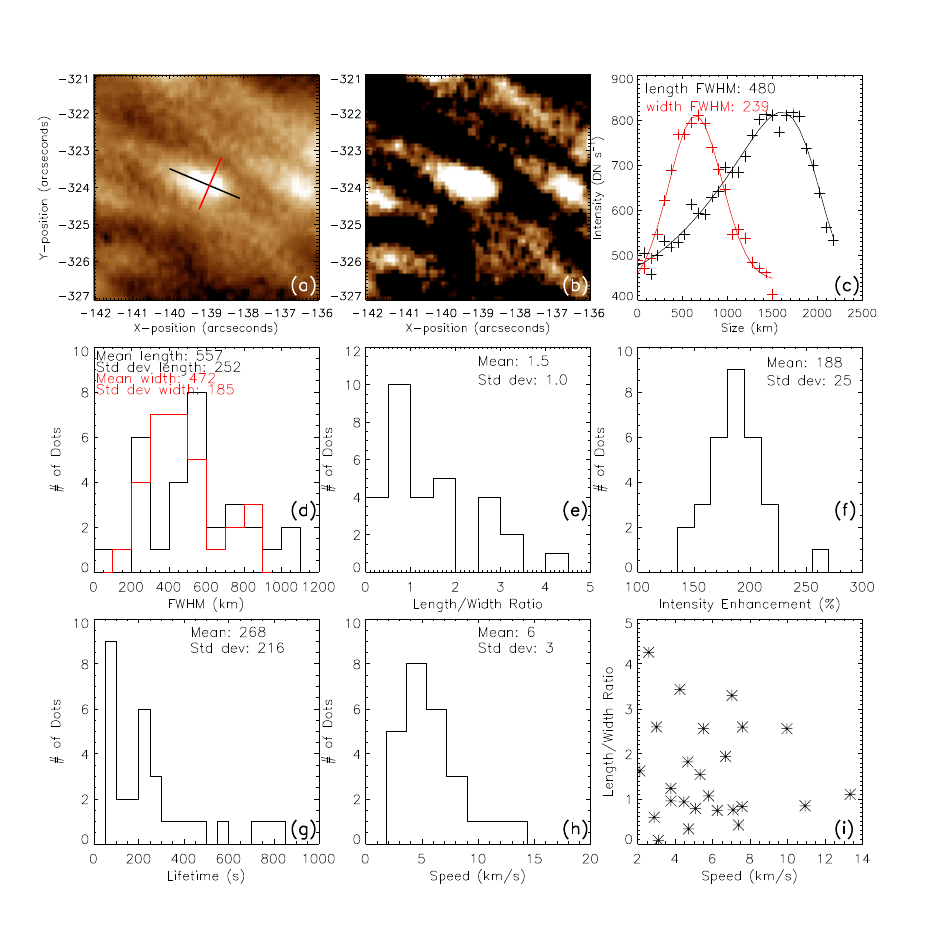}  
          \caption{Characteristics of Hi-C penumbral BDs. (a) Two cuts are used to measure the size of each BD. Black is for measurement of length (in the direction along the penumbral striations) and red is for measurement of width (perpendicular to length). (b) Same BD as in a), but enhanced image shown. (c) Example of BD size measurement; intensity is shown as a function of both length and width. A Gaussian fit is made to each cut and the FWHM is taken as the length/width. (d) Histogram of all length and width measurements. (e) Histogram of the ratio of length to width. (f) Histogram of how much brighter a BD is than its surrounding background. (g) Histogram of lifetimes of the BDs. (h) Histogram of the speed of the BDs. (i) Scatter plot of length/width ratio as a function of BD speed. }
     \label{f3}
\end{figure*}

Figure \ref{f3} shows histograms of the size, intensity, lifetime, and speed measurements of the 30 BDs that we selected in this sunspot penumbra. This figure is similar to Figure 3 of \cite{tian14} and we replicate their measurement methods closely in order to make the best comparisons between the properties of IRIS BDs and our BDs. Figures \ref{f3}a,b,c depict how the size was determined. Noticing the slightly non-circular shape of the BDs and in being consistent with the techniques used by \cite{tian14}, we measured length and width for each BD. The measurement was taken when the BD was the brightest. Figure \ref{f3}a shows one example BD in the normal intensity image. The length, measured along the black line, is in the local direction of the penumbral striation. The width is measured along the red line, perpendicular to the length cut. Figure \ref{f3}b shows the same dot as in \ref{f3}a, but from the enhanced image, which is used to determine the direction and placement of the dots.

In Figure \ref{f3}c, we show plots of intensity along each cut fitted with a Gaussian. Please note that a non-constant background is used when fitting the Gaussian. The size of the dot (length and width) is the full-width at half-maximum of the Gaussian. Figure \ref{f3}d displays the distribution of 30 BD lengths and widths; the length has a mean of 557 km and standard deviation of 252 km while the width has a mean of 472 km and a standard deviation of 185 km. Figure \ref{f3}e is the length-to-width ratio, showing that about half of the BDs are longer than they are wide. On average, BDs are not significantly larger in their direction of motion than in the orthogonal direction. Figure \ref{f3}f shows the intensity enhancement of each BD, measured as the brightness of the brightest part of the dot divided by the background brightness, which is estimated by averaging intensities of many pixels surrounding the BD. Average BDs are just under twice brighter than their surroundings.

The histogram in Figure \ref{f3}g represents the lifetime of the BDs, tracked manually and supplemented with AIA 193 \AA\ data for the longer-lasting BDs. Most of them last 1--5 minutes, but several survive for 12--14 minutes. The longer-lasting BDs are normally the bright, large, stationary ones and are closer to the sunspot umbra. 

Figure \ref{f3}h displays a histogram of the speed of the BDs. To calculate the speed of the BDs, we track the center of the BD as it moves. The tracking is done as follows. The initial location of the center of the BD is recorded and then the image is advanced in time. The location is then recorded again. The velocity is calculated by dividing the distance the center of the BD moved by the time between recordings. The image was then advanced again and this process is repeated until the BD disappears or merges with another BD. Because the lifetime of the BDs is variable, the time between recordings varies, but as a general rule, the number of recordings is maximized (minimizing the time between each recording) but there are at least 11 seconds between each recording. The average velocity of a BD is the average of these multiple velocity measurements. The speed was measured for only 25 BDs because there were not enough data points for five of the BDs. Figure \ref{f3}i is a scatterplot between the speed and the ratio of length-to-width of the BDs. We plot this figure in an attempt to find a correlation between how long a BD is relative to how fast it is moving (the thought being that faster BDs might be elongated in the radial direction), but evidently there is no clear relationship; there is perhaps a slight tendency for faster BDs to have less elongation.  

\begin{figure*}[h]
     \centering
     \includegraphics[width=0.37\textwidth]{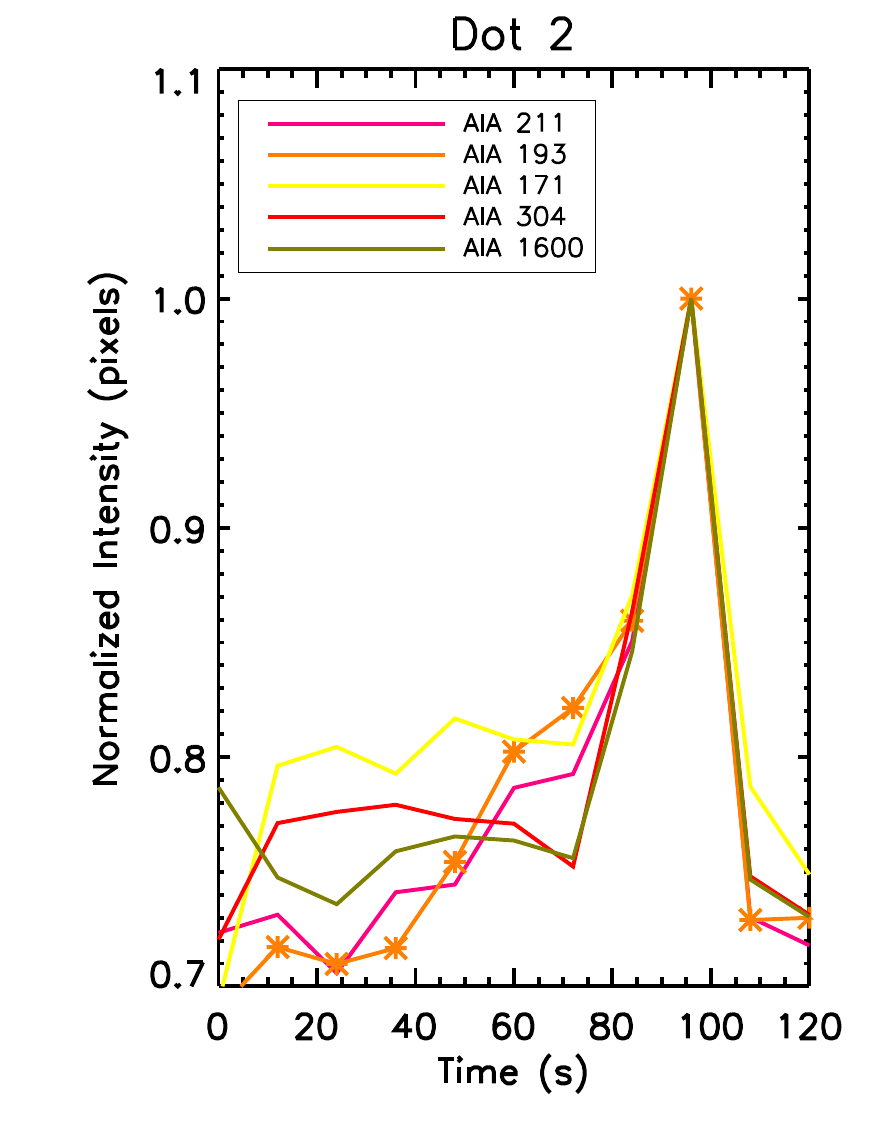}  
     \includegraphics[width=0.37\textwidth]{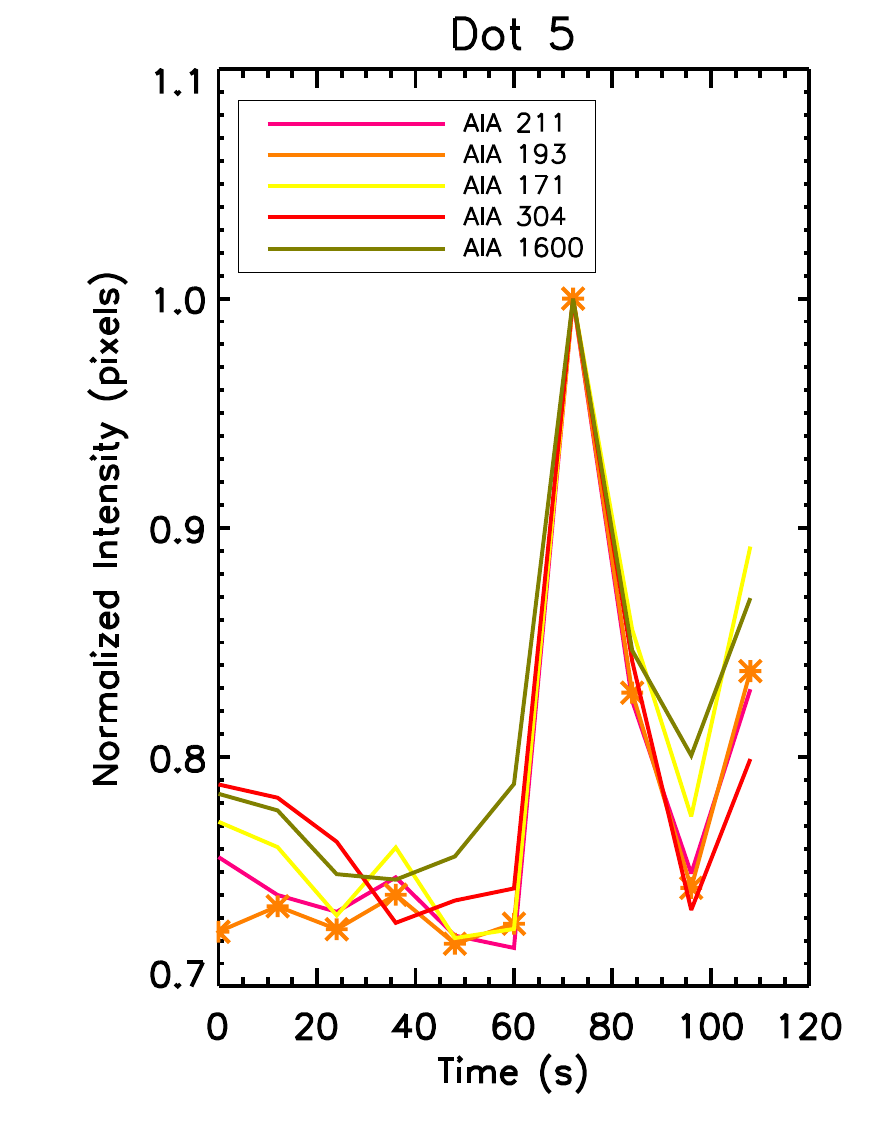}
     \includegraphics[width=0.37\textwidth]{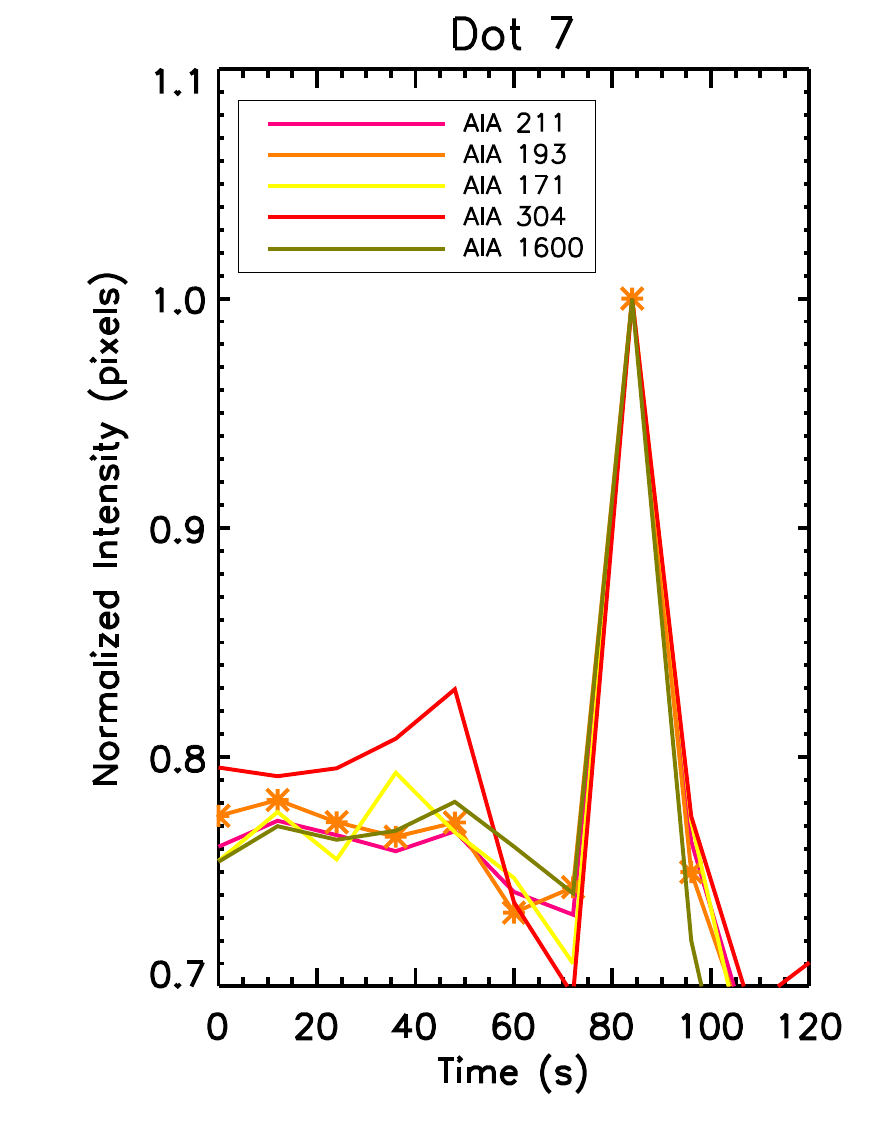}
     \includegraphics[width=0.37\textwidth]{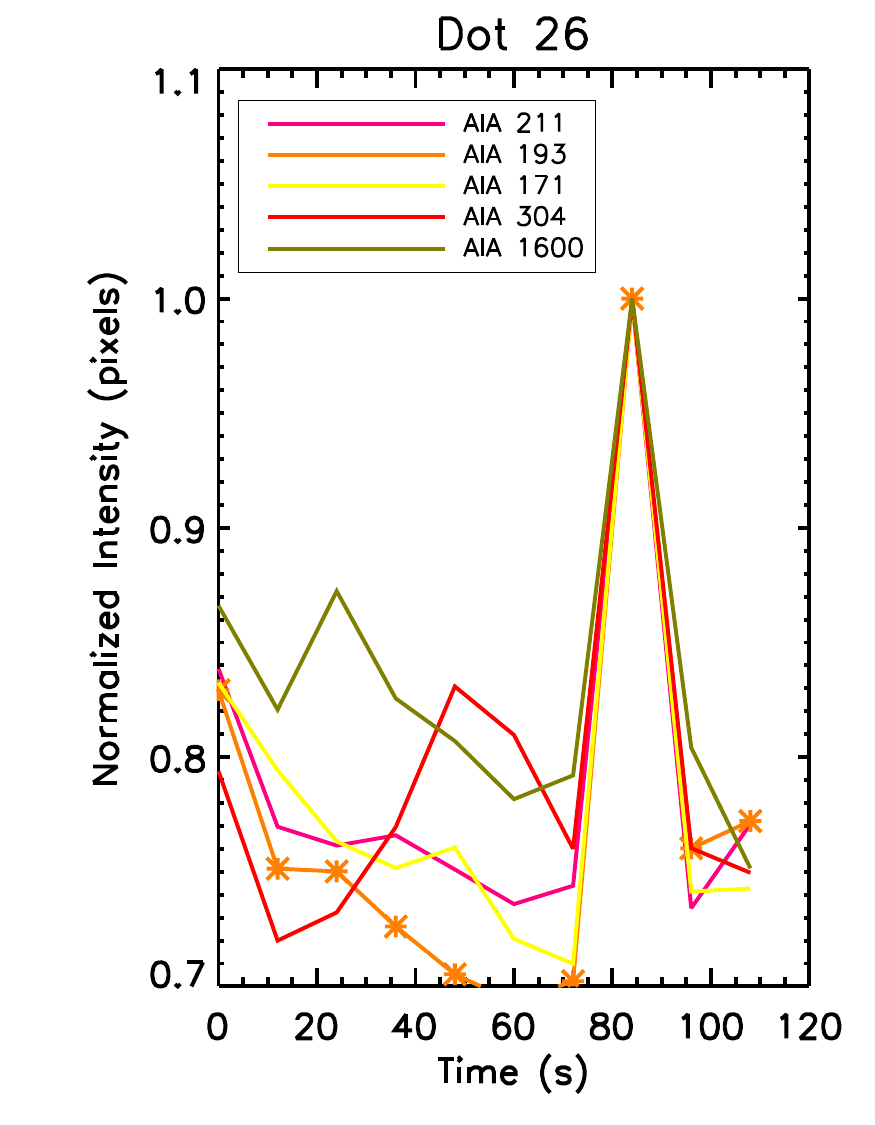}
        \caption{Normalized lightcurves for four example BDs; from a set of five AIA bands. Each set of lightcurves is from tracking the brightest part of the BD as it drifts laterally during its peak brightening. As denoted in each panel by different colored lines, orange is for 193, pink is for 211, yellow is for 171, green is for 1600, and red is for 304 \AA. Please note that the time on x-axis is based on the AIA 193 channel, the closest available data in time for other channels is taken at each time step. Therefore a few seconds difference in the observation of different channels is not seen. Asterisks on 193 lightcurves mark data points.}
     \label{f4}
\end{figure*}

To investigate the formation mechanism and temperature of these BDs, we obtained their lightcurves from five AIA channels. Figure \ref{f4} shows the lightcurves of four of our BDs. These lightcurves are made by tracking the BD from when it first begins to brighten, through peak intensity, and ending when it fades from view or gets dim enough. To measure the intensity, we first find the target BD using the Hi-C 193 \AA\ images and then match the location to the AIA 193 \AA\ images. We center a circle on the BD's brightest part and measure the intensity inside. The frame is advanced and the intensity is measured for the new time, again with the circle centered on the brightest part of the BD. In this way, we track the BD's peak brightness over its lifetime or over its brightest peak during its lifetime. This intensity is then normalized to the highest intensity over the  period. 


We use four AIA wavelengths in addition to the 193 \AA: 211, 171, 1600, and 304 \AA. The AIA 171, 193, and 211 channels are typically coronal, as their response functions peak at temperatures of roughly 1, 1.5, and 2.5 MK, respectively \citep{leme12}. However, all of these channels have cooler-TR lines in the passband and hence have some response in the cooler-TR temperature range \citep[e.g.,][]{odwy10}. The 304 and 1600 channels both respond to cooler-TR emission. If the BDs have cooler-TR temperature then these different wavelength channels, which all have some sensitivity to emission from plasma at TR temperature, will peak at the same time \citep{wine13}. If the BDs are heated to (and cooling from) coronal temperatures, then the different channels will not peak at the same time. Figure \ref{f4} shows the lightcurves of a few representative BDs, showing that the different wavelengths do peak at the same time.

These lightcurves indicate that the BD plasma is at cooler-TR temperatures ($1-4 \times 10^5$ K); its appearance in AIA coronal channels is due to leakage of low temperature emission into the predominantly coronal channels. Note that the lightcurves before and/or after the peak brightness need not necessarily be in phase.

After determining the characteristics of the BDs that we see with Hi-C, we compare them to the BDs seen by \cite{tian14} using the IRIS telescope. Table 1 compares our measured BD values with those found in \cite{tian14}. All listed values are averages.

\begin{table}\label{t1}
\begin{center}
\caption{Comparison of penumbral BD average quantities measured in \cite{tian14} and in this work.}
\begin{tabular}{| l | c | r |}
\hline
& Tian et al. & This research  
\\
\hline
Length (km) & 439 & 558
\\
\hline
Width (km) & 352 & 472
\\ 
\hline
Intensity Enhancement (\%) & 319 & 188
\\
\hline
Lifetime (s) & 41 & 268
\\
\hline
Speed (km/s) &22 & 6
\\
\hline
\end{tabular}
\end{center}
\end{table}

\cite{tian14} did not calculate average speed because they measured dot velocities, i.e., they included the direction the dot moves in their data. In this work, we measure the speed of the BD as the distance traveled by the center of the BD over time, without reference to the direction of travel. 
We deduce from Table 1 that our BDs are, on average, larger, longer-lived, and slower than the dots seen in the TR with IRIS. These results and our finding from the lightcurves that the Hi-C BDs have cooler-TR temperature implies that Hi-C detects only the brighter BDs seen by IRIS. The smaller and dimmer BDs are probably masked by the \FeXII\ emission of the overlying corona.

Based on our analysis, there are several different mechanisms that could be the cause of these BDs. Because the BDs are at the feet of coronal loops, cooled plasma that flows down along the loops could hit the lower, denser atmosphere and create shocks that locally heat the plasma and cause an increase in intensity; this requires supersonic downflow ($\sim$200 km $s^{-1}$) \citep{klei14}. The heating could also be due to the intricate, interlaced magnetic fields of the penumbra repeatedly reconnecting; the radially inward/outward motion of BDs might result from sliding reconnection between the two inclined penumbral field components (spines and bulk of penumbral filaments). Also using Hi-C data, \cite{regn14} found EUV bright dots at the edges of active regions which have a characteristic diameter of 680 km and a duration of 25 s; they proposed impulsive energy release in the coronal loops \citep{parker88} rooted in the dots being responsible for those dots. Ubiquitous chromospheric bright grains have also been observed in coronal holes and the quiet Sun, also on the order of 1\arcsec in diameter and lasting 1.5--2.5 minutes \citep[e.g.,][]{mart15}. 

Running penumbral waves \citep{ziri72,moor73,bloo07}, which are driven by photospheric p-mode oscillations, could perhaps be responsible for the oscillatory motion of some of the BDs. These waves could cause the BDs to move along field lines rooted in the penumbra. However, there are very slow moving and stationary BDs present near the BDs that oscillate; this casts doubt on running penumbral waves as the cause of all of our BDs. 

The slowest moving or stationary BDs might be at the heads of penumbral filaments \citep{tiw13}, also known as penumbral grains \citep{mull73,rimm06}. Those BDs might then be the result of opposite polarity fields at the sides of filaments reconnecting with the surrounding spines. This mechanism would also create penumbral jets \citep{kats07}, as proposed by \cite{tiw16}. Thus, some penumbral jets and BDs perhaps have the same location and same cause. This proposal is similar to the one given by \cite{viss15} that some penumbral BDs are the heating signatures of penumbral jets. However, from Figure 2 of \cite{tiw16} we find evidence that most of the Hi-C penumbral BDs do not coincide with penumbral jets. As pointed out by \cite{tian14}, some of the BDs concievably could be produced by moving magnetic features \citep{zhan03,ravi06,sain08}. 

\section{Conclusions}

We observe BDs in the penumbra of a sunspot using 193 \AA\ data from the Hi-C. Their lateral motion is inward or outward from the sunspot center, following the direction of penumbral filaments. We compare our BDs to those found by \cite{tian14} in the TR using the IRIS telescope. Our BDs are 300--800 km long and 300--600 km wide, last 1--5 minutes or longer, are double the intensity of the surrounding regions, and move 3--15 km s$^{-1}$. In general, they are larger, longer-lived, and slower than the IRIS BDs. Based on the lightcurves obtained using supplemental AIA data, most of the plasma in our BDs is probably at cooler-TR temperature, but the data is widely varied and subject to resolution and temporal constraints. The cause of these BDs is an open question. They are conceivably a consequence of plasma downflow impacting the TR or upper-chromosphere density, or of reconnection between two inclined components of penumbral magnetic field (more vertical spine field and more horizontal field in the bulk of penumbral filaments), perhaps sometimes triggered by running penumbral waves. The Hi-C BDs belong to the class of TR penumbral BDs found in IRIS data by \cite{tian14}, and have possible counterparts at the edges of active regions in the corona and/or TR \citep{regn14}.

\acknowledgments
This work is supported by NSF Grant No. AGS-1157027, a Research Experience for Undergraduates grant. These BDs were reported in 2014 at the LWS Science Meeting and at the AGU meeting on November 2-6 and December 15-19, respectively. Thanks goes to Hui Tian for helpful discussion on BDs and the methods of measurement. SKT is supported by an appointment to the NASA Postdoctoral Program at the NASA MSFC, administered by USRA through a contract with NASA. RLM and AW were supported by funding from the LWS TRT Program of the Heliophysics Division of NASA's SMD. SLS is supported through the Hinode project office as part of NASA SMD's Solar Terrestrial Probes Program.


\end{document}